\documentclass[useAMS,usenatbib]{mn2e}
\input epsf
\usepackage{graphicx}
\sloppypar

\title[Thermal diffusion]{Thermal diffusion in the IGM of clusters of galaxies}
\author[P. Shtykovskiy and M. Gilfanov]{P. Shtykovskiy$^{1,2}$
\thanks{E-mail: pavel@iki.rssi.ru; gilfanov@mpa-garching.mpg.de} 
and M. Gilfanov$^{2,1}$\footnotemark[1]\\
$^{1}$Space Research Institute, Russian Academy of Sciences, 
      Profsoyuznaya 84/32, 117997 Moscow, Russia\\
$^{2}$Max-Planck-Institute f\"ur Astrophysik, 
      Karl-Schwarzschild-Str. 1, D-85740 Garching bei M\"unchen, 
      Germany}
\begin{document}
 
\date{Accepted. Received; in original form}

\pagerange{\pageref{firstpage}--\pageref{lastpage}} \pubyear{2009}

\maketitle

\label{firstpage}

\begin{abstract}
We revisit the phenomenon of elements diffusion  in the intergalactic medium (IGM) 
in clusters of galaxies. The diffusion is driven by  gravity, 
concentration and temperature gradients. The latter cause thermal 
diffusion, which has been so far  ignored in IGM studies.
We consider the full problem based on the Burgers' equations and demonstrate that 
the temperature gradients present in clusters of galaxies may
successfully compete with gravity, evacuating metals from cooler regions.
Under the combined action of gravity 
and temperature gradients, complicated metallicity profiles with several  peaks and depressions  
may be formed.  For a typical cool core cluster, the thermal diffusion may significantly 
reduce and even reverse the gravitational sedimentation of metals, resulting in the depression 
in their  abundance in the core.
This  may have implications for diagnostics of the low temperature plasma in the centers of
 clusters of galaxies.

\end{abstract}


\section{Introduction}
\label{sec:intro}

Clusters of galaxies are the largest gravitationally-bound objects
in the Universe with  thousands of galaxies residing in a 
dark matter-dominated potential.
Hot intergalactic medium (IGM), constituting the majority 
of the baryonic matter in galaxy clusters, is a natural laboratory 
for plasma physics displaying  a wide range of phenomena.
Despite seemingly simple nature of galaxy clusters,    understanding
of   physical processes in hot plasma of IGM is far from being complete.

Of particular importance is the heavy elements distribution.
Indeed, on the one hand it affects significantly the outcome of 
cosmological measurements via the mean molecular weight and thus the
 derived  mass of the clusters \citep[e.g.][]{qin00,chuzhoy03} 
and the baryonic density \citep[e.g.][]{ettori06}.
On the other hand the X-ray spectrum of gas at low temperatures,
T$\la3\times10^{7}$~K, is dominated by the emission lines,  most important of which are 
associated with iron.    
The diagnostics of low temperature plasma  in clusters of galaxies
is based primarily on line emission from heavy elements and the distribution of the latter  
 may have direct implications for various phenomena, 
including the cooling flow problem 
\citep{fabian94}. 

Despite the importance of the heavy element distribution, 
 flat metallicity profiles are often assumed.
This assumption may turn out to be too crude 
in some cases. Indeed, the starting point for the metallicity profile is
the primordial abundance.   
It is, however, quickly modified by a  number of 
competing processes such as metals injection by supernovae, mixing and diffusion.
Although the effects of the  latter process are rather controversial and its relevance to
 IGM in clusters of galaxies is still under debate, 
the scenario where the diffusion plays a significant role in forming the metallicity profile
does not seem to be entirely unrealistic.

The diffusion of heavy elements in the intracluster medium 
has been considered by e.g. \citet{fabian77,gilfanov84,chuzhoy03}. 
Most studies focused on the  gravitational
separation (gravitational sedimentation)
and found that it may result in significant changes of the 
gas composition throughout the cluster.
Given the presence of temperature gradients in IGM,
an additional type of diffusion may play a role, namely thermal diffusion.
Although the importance of the latter in physics in general and in stellar 
physics in particular has been appreciated long ago  \citep[e.g.][]{thoul94},
it was largely ignored in the case of ISM in galaxies and IGM in cluster of galaxies.
The exceptions are studies by \citet{chuzhoy04} who solved the full problem, however 
considered the case of isothermal cluster and 
\citet{peng08} who mentioned the potential relevance of the thermal diffusion effects
in galaxy clusters.
At first glance, simplified approach ignoring thermal diffusion may
seem to be reasonable since the temperature variations
in galaxy clusters are rather small. It turns out, however, that 
even moderate temperature 
gradient may be sufficient to compete the gravitational force.
The consideration of the full problem including the effects of thermal 
diffusion is the subject of present paper.

The paper is structured as follows.
In section~\ref{sec:theory} we discuss the nature of 
the metallicity profiles in galaxy clusters. 
In section~\ref{sec:experiment} we describe the 
basic equations, numerical scheme
and simulate the diffusion-driven evolution of 
a model galaxy cluster.
Finally, in section~\ref{sec:discussion} we
discuss the obtained results.

\section{Metallicity profiles in galaxy clusters}
\label{sec:theory}

Although the observed metallicity profiles  are known to vary from cluster to cluster,  
they do exhibit some similarities in their  behaviour
\citep[e.g.][]{vikhlinin05,sanderson09}.
Typically, in cool core (CC) clusters  the heavy element abundance rises from
the cluster outskirts to the core. 
At the same time in the central parts, $r\la10-20$~kpc, a fraction of CC  clusters show 
a depression in metallicity, whose nature and even reality is still under
 debate \citep[e.g.][]{werner09}.

The metallicity profiles in galaxy clusters form as a result of interplay
between different processes such as
injection of metals from the supernovae-driven galactic winds, turbulent mixing and diffusion.
The importance of the latter in plasma with turbulent magnetic field 
has been debated for a long time \citep[e.g.][]{bakunin05,lazarian07}. 
As it is well known, a magnetic field hinders the diffusion across the field lines.
However, given the chaotic character of the field in a turbulent 
medium, the efficiency of transport processes may be high enough 
to make the diffusion and thermal conduction important \citep{narayan01}.
In the following we assume that this is indeed the case.
\medskip 

The diffusion is driven by the gravity, 
temperature and concentration gradients.
Below we consider each in details.

\begin{enumerate}
\item {\it Gravitational sedimentation.}
The equilibrium distribution of heavy elements, n$_s\propto e^{-m_sgh/kT}$, 
is more concentrated 
towards the cluster's center than that of the hydrogen.
Thus, due to the gravitational force, heavy elements tend to sink 
towards the cluster center, while light elements rise.

\item {\it Thermal diffusion.} 
Thermal diffusion was discovered theoretically by Enskog  
and Chapman in the 1911-1917 \citep[e.g.][]{chapman52}.
A clear physical explanation of the effect has been given significantly 
later by \citet{monchick67}, who demonstrated that thermal diffusion results 
from the dependence of collisional frequency on the velocity of particles.
In the gas composed of two elements,  the effect of thermal diffusion is to 
evacuate the more massive and more highly charged species from the colder 
region of the gas.
In the case of a gas with three species -- the abundant light ions,
electrons and a small admixture 
of heavy ions with charge Z$_s$ and atomic masses A$_s$, 
the equilibrium distribution of the heavy element 
may be obtained analytically \citep{burgers69}:

\begin{equation}
n_s(r)\propto T^{\alpha Z_s^2}\exp\left[-\frac{(A_s-0.5\,Z_s)\,m_p}{k_B}\int \frac{g(r)}{T(r)}\,dr\right], 
\label{eq:equildistr}
\end{equation}
where alpha is constant of the order of 3.
This formula demonstrates the importance of thermal diffusion 
for heavy elements. Indeed, substituting Z$_S$=26  for the iron, we obtain
$\ga2000$ for the power of T in the first multiplier.
In practice, however, the time required to reach 
the equilibrium distribution may easily exceed the Hubble time and
such a steep abindance profile is not formed.

\item {\it Concentration gradients.}
The diffusion  driven by concentration gradients 
works to restore the uniform abundance profile.
\end{enumerate}

Another process affecting the heavy elements distribution is turbulent mixing 
which tends to wash out the metallicity gradients. 
The effects of mixing depend on the spatial scales over which 
the turbulence is operating and on the chaotic velocities spectrum.
Mixing may also be driven by dynamical effects such as ``sloshing''
\citep{ascasibar06}. 

\medskip 
In the next section we construct a model cluster of galaxies and 
calculate the evolution of its metallicity profile  due to
diffusion of elements.
Our goal is to consider the full problem of diffusion for an idealized cluster in order to demonstrate the effects of 
heavy elements transport in galaxy clusters, with particular emphasis on the
 modifications brought about by the thermal diffusion. 
Construction of a fully self-consistent model is beyond the scope 
of this paper, therefore we make several simplifying assumptions, namely:

\begin{enumerate} 

\item We consider the idealized case of a static cluster assuming 
that it's mass remains constant throughout the duration of simulations.
In real case the cluster mass continiously grows via mergers and accretion
of DM halos \citep[e.g.][]{cohn05}. 
Since we are mostly interested in cool core clusters which 
are (most likely) not affected significantly by mergers in the past few 
Gyrs \citep[e.g.][]{santos08,poole08}, this should not compromise  our results
significantly. 

\item We assume that there is an exact balance between heating and cooling of IGM 
 which maintains  the temperature profile constant throughout the simulations.
This implies that the IGM heating via the thermal conduction 
(which accompanies diffusion as these are the two sides of one phenomenon) 
and due to action of the central AGN  equal the radiative cooling. 
As shown by \citet{guo08}, a stable configuration of this kind may indeed
exist.

\item The IGM enrichment 
due to the continuous injection of metals from galaxies is ignored.

\item The effect of mixing on the metallicity profile of the cluster is ignored.

\end{enumerate}
The possible effect of the latter two processes is discussed in the section \ref{sec:mixing}.

\section{Numerical experiment}
\label{sec:experiment}

We consider a spherically-symmetric cluster of  galaxies with multicomponent plasma 
residing in a dark matter dominated gravitational potential.
The latter is assumed to follow the NFW profile \citep{navarro97}.
The cluster is assumed to be initially in the hydrostatic equilibrium.

\subsection{Basic equations}
\label{sec:equations}

We consider $M$ species ($M-1$ ions plus electrons) with 
concentrations $n_s$, mean velocities $u_s$, 
atomic masses $A_s$ and charges $Z_s$. All species have the same temperature 
$T_s\equiv T$. The diffusion velocities $w_s$ are defined as:
\begin{equation}
w_s\equiv u_s-u, 
\label{eq:diffvelocity}
\end{equation}
where u is the mean fluid velocity:
\begin{equation}
u=\frac{\sum_s n_sm_su_s}{\sum_s n_sm_s}.
\label{eq:meanfluidvel}
\end{equation}

The hydrodynamics of multicomponent gases is governed by Burgers' equations
\citep{burgers69}.
The Burgers' equations for the gas in the hydrostatic
equilibrium (the mass, momentum and energy conservation equations) 
are \citep[][]{burgers69,thoul94}:
\begin{equation}
\frac{\partial{n_s}}{\partial{t}}+\frac{1}{r^2}\frac{\partial}{\partial{r}}[r^2n_s\,(w_s+u)]=0,
\label{eq:continuity}
\end{equation}
\begin{eqnarray}
\frac{d(n_sk_BT)}{dr}+n_sm_sg-n_sZ_seE=\sum_{t\neq
  s}K_{st}[(w_t-w_S)+
\label{eq:euler}
\\
+0.6(x_{st}r_s-y_{st}r_t)],
\nonumber
\end{eqnarray}
and 
\begin{eqnarray}
\frac{5}{2}n_sk_B\frac{dT}{dr}=\sum_{t\neq
  s}K_{st}\{\frac{3}{2}x_{st}(w_s-w_t)-
\label{eq:energy}
\\
-y_{st}[1.6x_{st}(r_s+r_t)+Y_{st}r_s-4.3x_{st}r_t]\}-0.8K_{ss}r_s.
\nonumber
\end{eqnarray}
Here $g=[GM(r)/r^2]$ is the gravitational acceleration {\it modulus}, $E$ is the
radial electric field, $x_{st}=\mu_{st}/m_s$, $y_{st}=\mu_{st}/m_t$ with 
$\mu_{st}=m_sm_t/(m_s+m_t)$ and $Y_{st}=3y_{st}+1.3x_{st}m_t/m_s$. 
Coefficients in equations correspond to pure Coulomb potential
with a long-range cutoff at the Debye length \citep{thoul94}.
The quantities $r_s$ are  the so-called residual heat flow vectors introduced by
 \citet{burgers69}, playing the role of additional independent variables.
The thermal diffusion arises due to the interference of 
equation~\ref{eq:euler} and the energy equation~\ref{eq:energy} 
via the $r_s$.

The friction coefficient between species s an t is:
\begin{equation}
K_{st}=(2/3)\mu_{st}(2k_BT/\mu_{st})^{1/2}n_sn_t\sigma_{st},  
\label{eq:frictioncoeff}
\end{equation}
where the cross section between particles s and t is:
\begin{equation}
\sigma_{st}=2\sqrt{\pi}e^4Z_s^2Z_t^2(k_BT)^{-2}\ln\Lambda_{st}.
\label{eq:crosssection}
\end{equation}
We assume the Coulomb logarithm to be $\ln\Lambda_{st}=40$ everywhere.

In addition, charge neutrality and the absence of currents should be satisfied
 throughout the plasma. The former is not imposed explicitly in our code and is 
used to control the solution, while the latter is used to close the system of Burgers' 
equations (see below).

\subsection{Simulations}
\label{sec:numhydro}

The galaxy cluster is assumed to be  initially in the hydrostatic
equilibrium, i.e. there is no net mass flow, $u=0$.
However, once diffusion starts to operate, the total pressure 
changes and the gas is pushed out of the equilibrium, $\nabla p+\rho g\neq0$.
Indeed, while heavy elements flow to the center of the cluster, 
hydrogen flows outwards and there is a net outflow of particles.
The gas  adjusts itself to the new equilibrium
configuration via the contraction on the hydrodynamical timescale.
Given the much smaller timescale of the latter as compared to the
diffusion timescale, the departure from equilibrium will be small
enough for the hydrostatic equilibrium form of Burgers' equations 
to be  valid throughout the lifetime of a cluster 
(see also sec.~\ref{sec:numhydrotests}).

In order to solve the problem we  use the following numerical scheme:
\begin{enumerate}
\item  Based on the concentrations of elements $n^{k-1}_s$ on k-1-th time step,
 solve Burgers' equations ~(\ref{eq:euler}) and
~(\ref{eq:energy}) for the diffusion velocities $w^{k-1}_s$.

\item Given the diffusion velocities $w^{k-1}_s$, the total mass flow
  velocity $u^{k-1}$ and $n^{k-1}_s$, solve the continuity equation 
for the concentrations on the next time slice $n^{k}_s$.

\item Solve the Euler equation
\begin{equation}
\frac{du}{dt}=-\frac{\nabla p}{\rho}-g
\label{eq:euler2}
\end{equation}
for the velocity $u^{k}$, restoring the hydrodynamic equilibrium in a
cluster, using $u^{k-1}$ and $n^{k-1}_s$.

\end{enumerate}

Below we discuss each step in detail.

\subsubsection{Diffusion velocity calculation.}
\label{sec:numhydro1}

Burgers' equations (eqs ~\ref{eq:euler} and
~\ref{eq:energy}) present a linear inverse problem of the form 
$A\times y=b$ on the diffusion velocities $w_s$, heat vectors of species
r$_s$ and electric field E.
The diffusion is driven by the gradients $\nabla T,\, \nabla n_s,\, \nabla p\, (g(r))$, 
with all these quantities residing on the r.h.s. b.
Given the M species we have 2M+1 unknowns and thus need 
2M+1 independent equations.
Burgers' equations give us only 2M-1 independent equations, because diffusion 
velocities appear in eqs.~\ref{eq:euler} and
~\ref{eq:energy} only in the form ($w_s-w_t$). The two additional conditions needed to close the system of 
Burger's equations are defined as follows:
\begin{equation}
\sum \rho_iw_i=0, 
\label{eq:addeq1}
\end{equation}
which follows from the definition of the diffusion velocity 
(eqs~\ref{eq:diffvelocity} and ~\ref{eq:meanfluidvel}).
Finally the absence of electric currents
 gives:
\begin{equation}
\sum Z_in_iu_i=0
\label{eq:addeq2}
\end{equation}

The resulting system of equations is solved for the $w_s$ and 
$r_s$ using the singular value decomposition (SVD) method \citep{press92}.

\subsubsection{Continuity and Euler equations.}
\label{sec:numhydro2}

The problem we are solving is a mixed advection-diffusion problem.
We have found that it can be solved with
sufficient accuracy using the combination of forward-time centered-space (FTCS) and Lax methods
\citep{toro99}, with the former applied to the continuity equation and the
latter to Euler equation.
The Lax method is known to introduce numerical diffusion;
however, we performed a number of tests and found that it is not crucial for our case 
(see Sec.~\ref{sec:numhydrotests}).

We use a homogenous grid spanning from 0 to 2 Mpc with step $dr=0.25$~kpc.
The timestep is chosen to be small enough to ensure the stability of the 
scheme, $dt=2-5\cdot10^{-5}$~Gyr.

Since we are mainly interested in the diffusion in the central region 
of the cluster, we are relatively free in the choice of the outer 
boundary condition. 
We assume that the galaxy cluster is immersed in the medium
with constant density, i.e. there is an infinite reservoir of gas which maintains  the 
gas density together with the element mass fractions 
at the outer boundary constant.
In calculating velocities at the outer boundary we use one-sided 
derivarives.

At the inner boundary $r=0$  the continuity equation 
 has a singular point.
Therefore we expand eq.~(\ref{eq:continuity}) in the 
vicinity of zero and obtain the folowing expression for the
concentration at the k-th time step:
\begin{equation}
n^k_{1, s}=n^{k-1}_{1,s}-3\times(dt/dr)n^{k-1}_{1,s}(u^{k-1}_2+w^{k-1}_{2,s}).
\label{eq:bc2}
\end{equation}
We also required that the density gradient should be equal 
to zero at $r=0$ and used a one-sided derivative for the velocity gradient
$\nabla u$.
In addition all  velocities should be zero at the inner point: $w_s=0$
and $u=0$.

\begin{figure*}
\begin{centering}\hbox{
\includegraphics[width=0.45\textwidth]{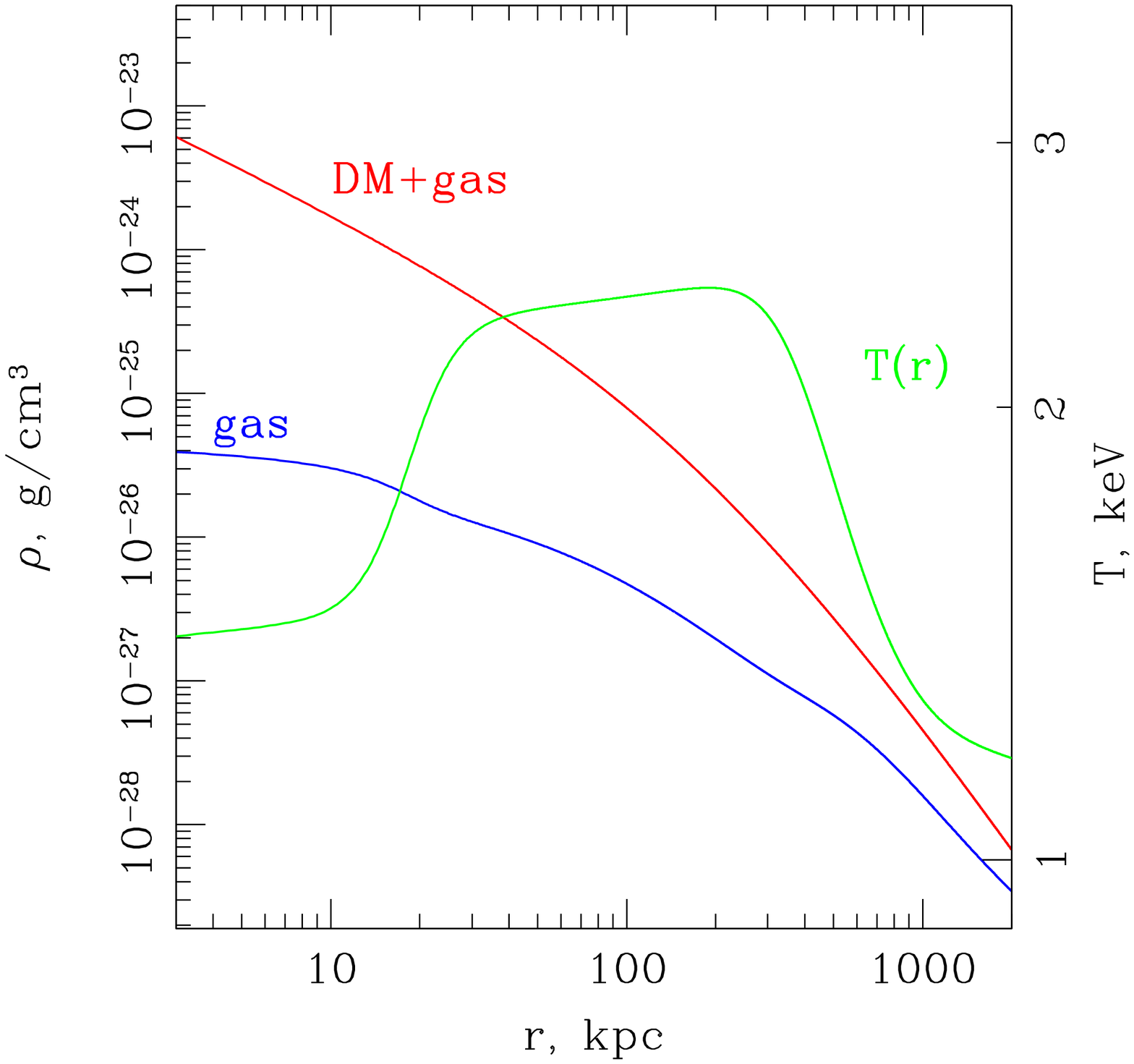}
\includegraphics[width=0.45\textwidth]{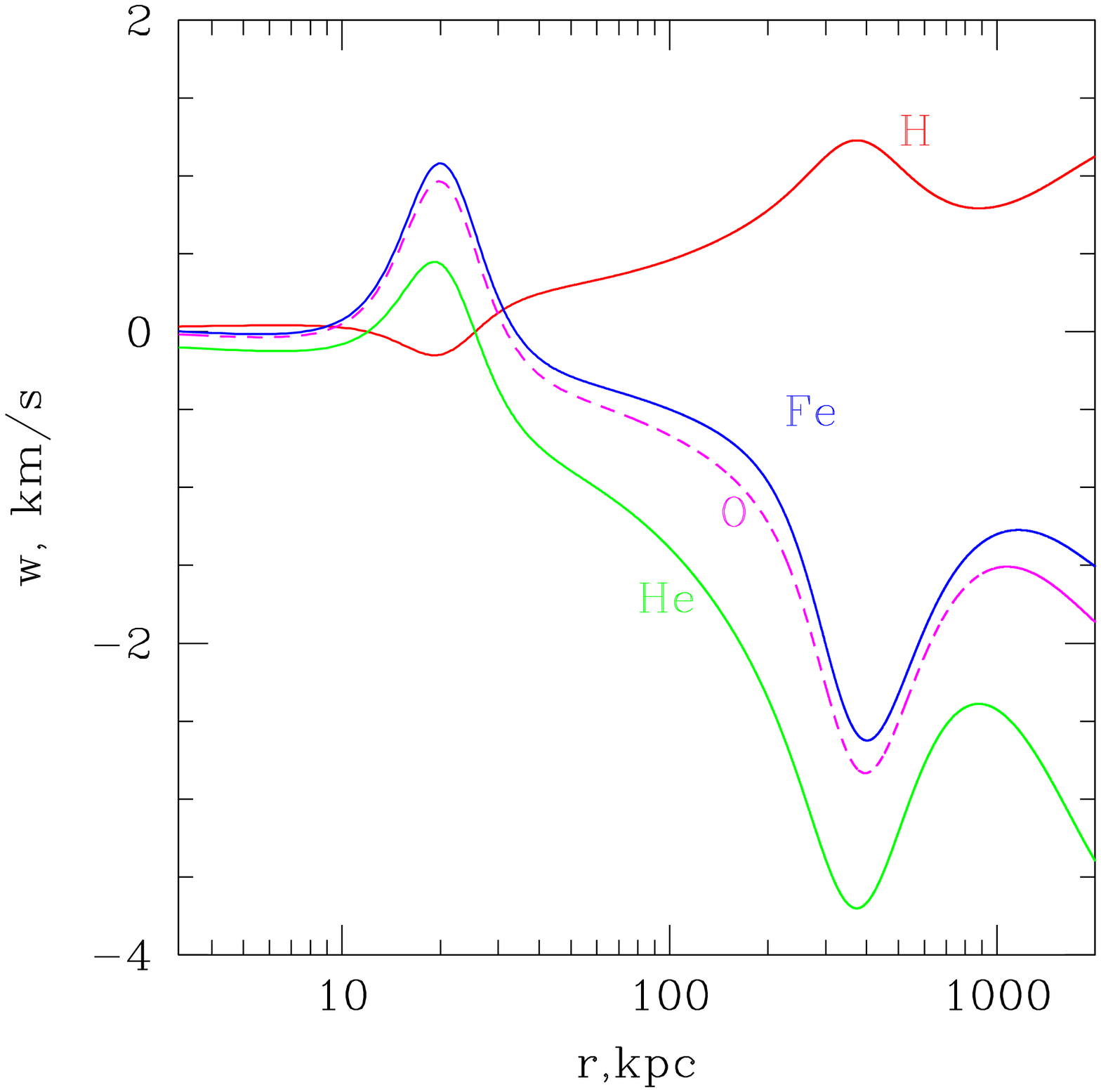}
}\end{centering}
\caption{{\em Left:}  Dark Matter and gas 
density profiles and temperature profile of the model
galaxy cluster.
{\em Right:} Corresponding initial diffusion velocity profiles for H (red), 
He (green), Fe (blue) and O (magenta, dashed line). Positive velocities correspond to outflow of particles, negative -- to inflow.
A positive peak at small radii and a negative one at the cluster outskirts are caused by the thermal diffusion and correspond 
to strong temperature gradients in the IGM (see the left panel).
}
\label{fig:profiles}
\end{figure*}

\begin{figure*}
\begin{centering}\hbox{
\includegraphics[width=0.45\textwidth]{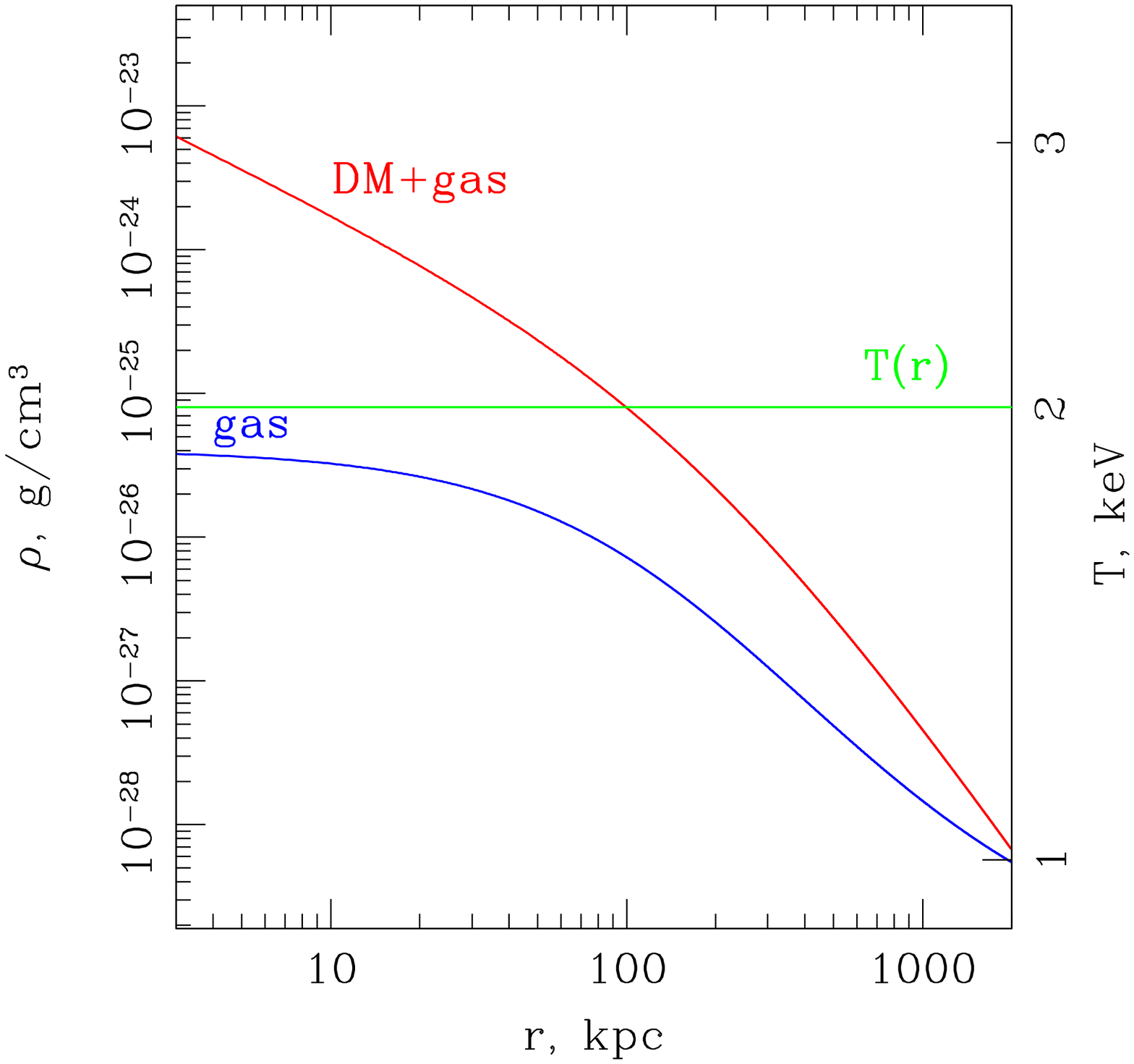}
\includegraphics[width=0.45\textwidth]{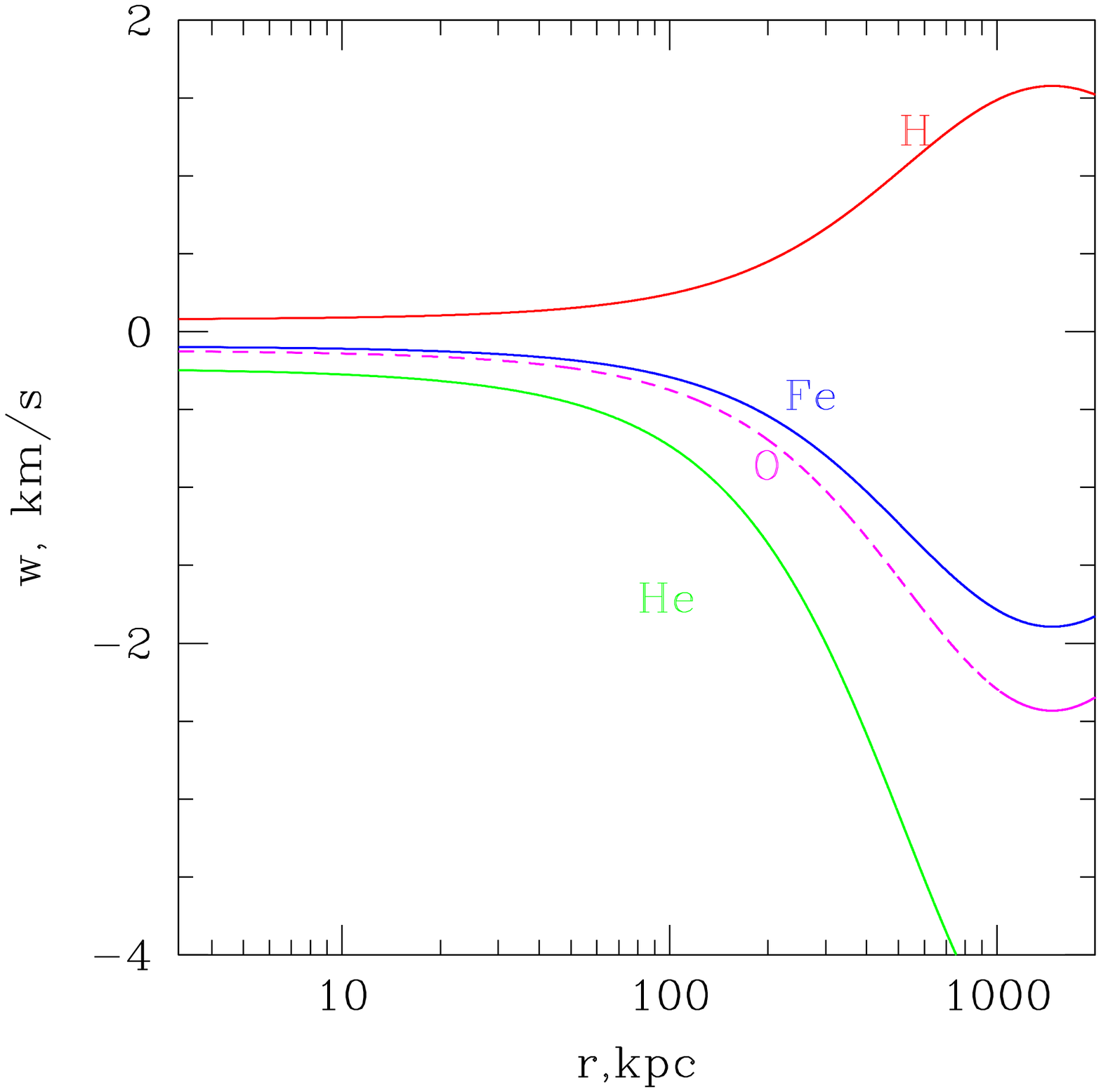}
}\end{centering}
\caption{Same as in Fig.\ref{fig:profiles} but for an isothermal cluster of galaxies.
}
\label{fig:profiles2}
\end{figure*}

\subsubsection{Code validation.}
\label{sec:numhydrotests}

We further performed a number of tests to check the validity of 
our code.

\begin{enumerate}

\item {\em Numerical diffusion.}
We obtained the solution replacing the Lax method with the
FTCS. The resulting code is free from numerical diffusion, however the FTCS-based scheme is less suited for our purpose due to  a well-known instability of FTCS for hydrodynamical problems. 
On the other hand since the hydrodynamical terms in our equations
appear only as a  result of a slow diffusion process, we were able to
obtain a stable solution
for the 7 Gyr  simulation. The solutions obtained using the combination 
of Lax+FTCS and FTCS+FTCS schemes are consistent with each other
within a few percent.

\item {\em Boundary conditions.}
We check that the solution is insensitive to  the boundary
conditions. For this purpose we
run our code for 7 Gyr  with different outer radii  and found no 
significant difference.
We also considered  the opaque
outer boundary, setting to zero all the velocities at the outer radius.
The  solution at $t=7$ Gyr in this case is also consistent with the one for the default
boundary conditions almost everywhere,  within a fraction of a percent.
Solutions obtained with different boundary conditions differ from each
 other only within $\sim 100$ kpc from the outer boundary. This dependence is 
unimportant and can be neglected.

\item {\em Hydrostatic equilibrium.}
We also checked that the
departures from the hydrostatic equilibrium are not too large, 
for the Burgers' equations to be valid, namely that 
$\vert\nabla p/\rho +g\vert\ll\vert\nabla p_s/\rho_s +g\vert$
throughout the duration of simulations.
This condition is satisfied with an accuracy of better than a 
fraction of a percent.

\item {\em Electro-neutrality.}
Since we do not use the electroneutrality condition 
explicitly in our code, we check that the final solution satisfies it.
We obtain that $\left| \sum Z_in_i/n_e-1\right|\la10^{-5}$.

\item {\em Comparison with an analytical solution.}
In order to check the implementation of the Burgers' equations in the code,
we consider a plasma composed of protons, 
electrons and a small admixture of iron in the hydrostatical equilibrium. 
Solving the Burgers' equations analytically we obtain
the coefficient $\alpha$ in the power of temperature in eq.~\ref{eq:equildistr},  $\alpha\approx3.1$, and compute the equilibrium distribution of iron.
For this distribution we  calculate the diffusion velocities using our code and verify that they equal zero as expected for an equilibrium distribution. We obtain, that $w_{\rm Fe}\ll10^{-2}$~km/s.

\end{enumerate}

\begin{figure*}
\begin{centering}\hbox{
\includegraphics[width=0.45\textwidth]{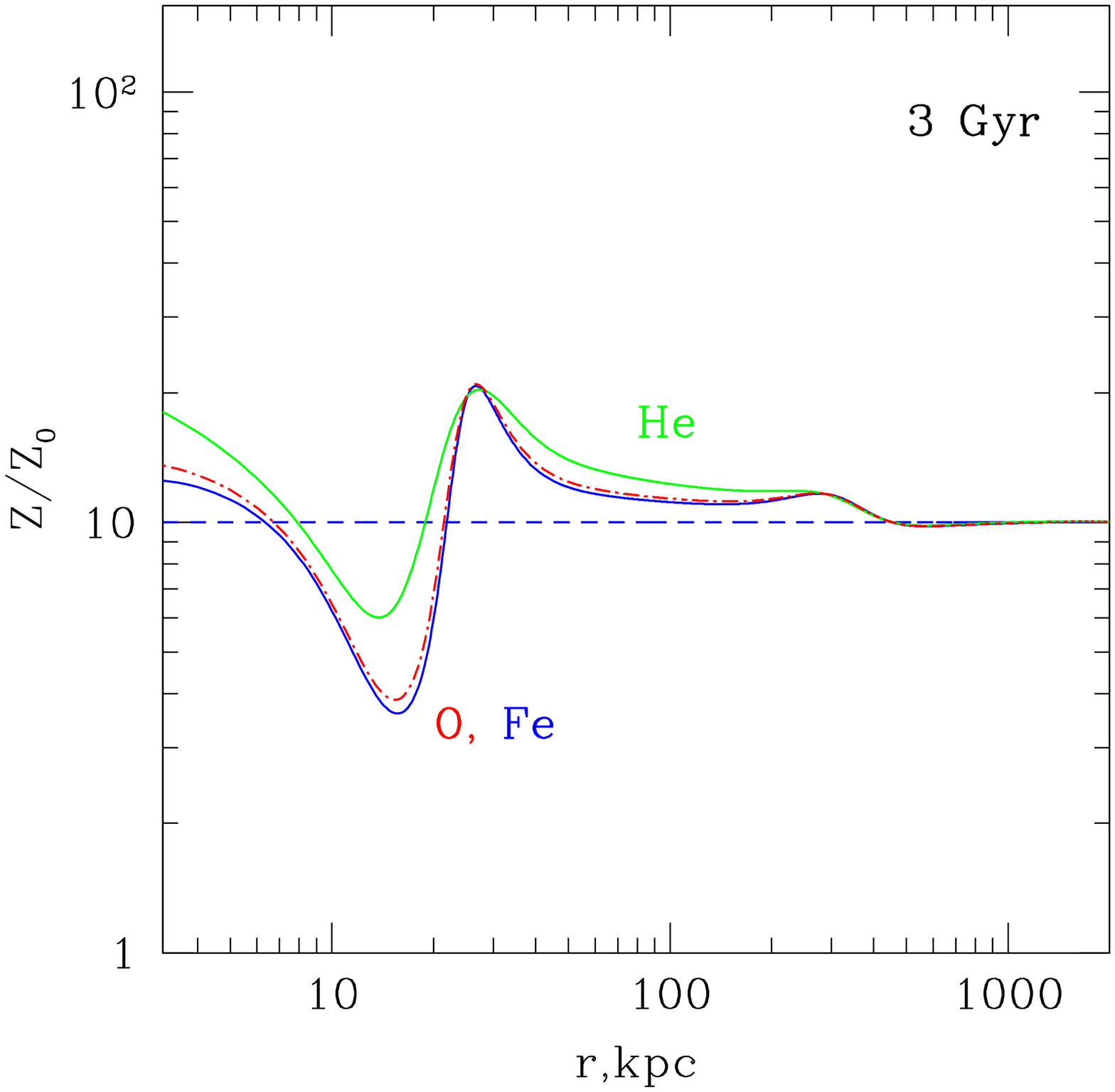}
\includegraphics[width=0.45\textwidth]{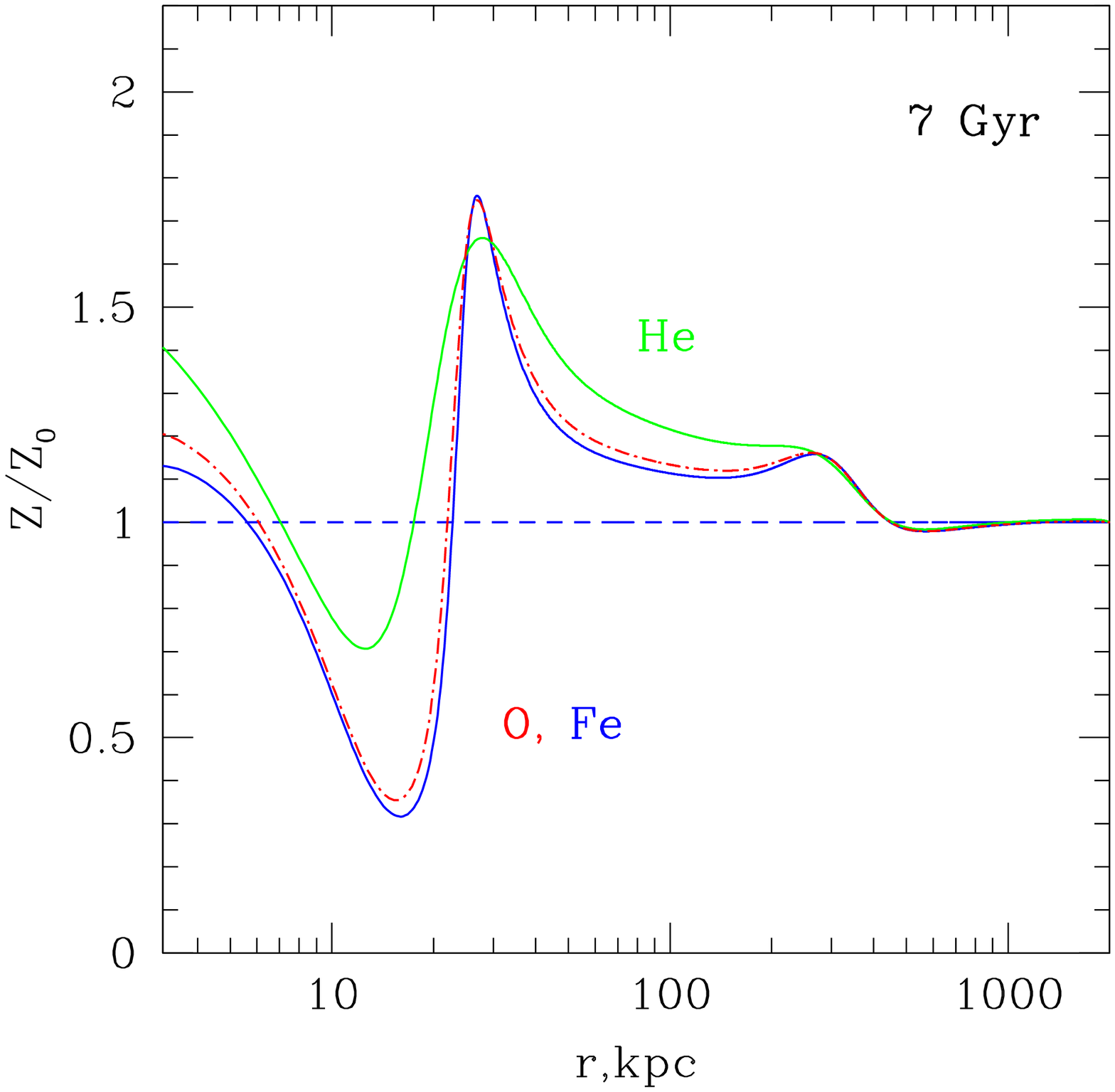}
}\end{centering}
\caption{Abundance profiles for Fe (blue solid line), O (red dot-dashed line) 
and He (green line) in the model cluster of galaxies from  Fig.~\ref{fig:profiles} after 3 Gyr (left) and 7 Gyr (right). 
}
\label{fig:metprofiles}
\end{figure*}

\begin{figure*}
\begin{centering}\hbox{
\includegraphics[width=0.45\textwidth]{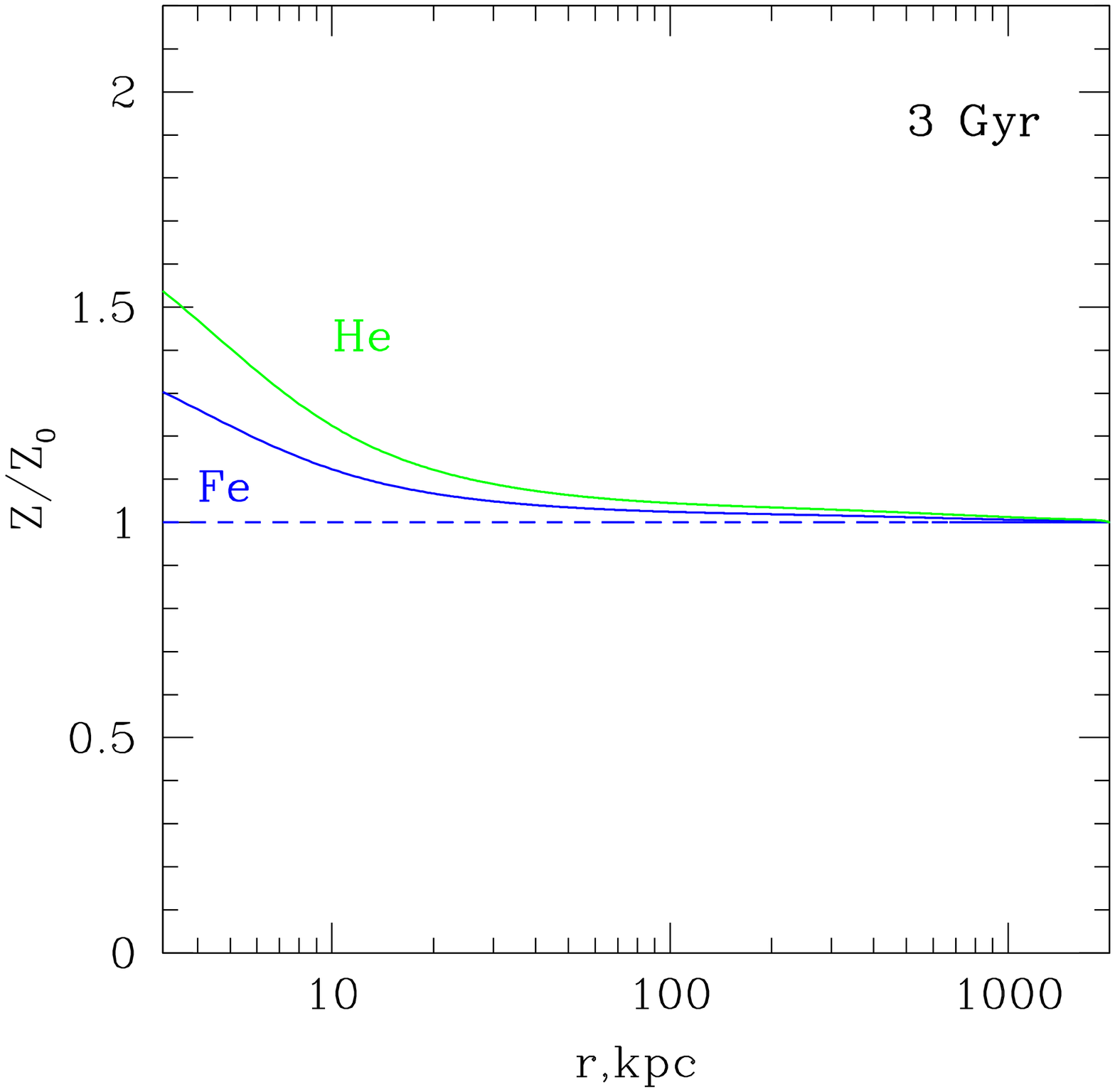}
\includegraphics[width=0.45\textwidth]{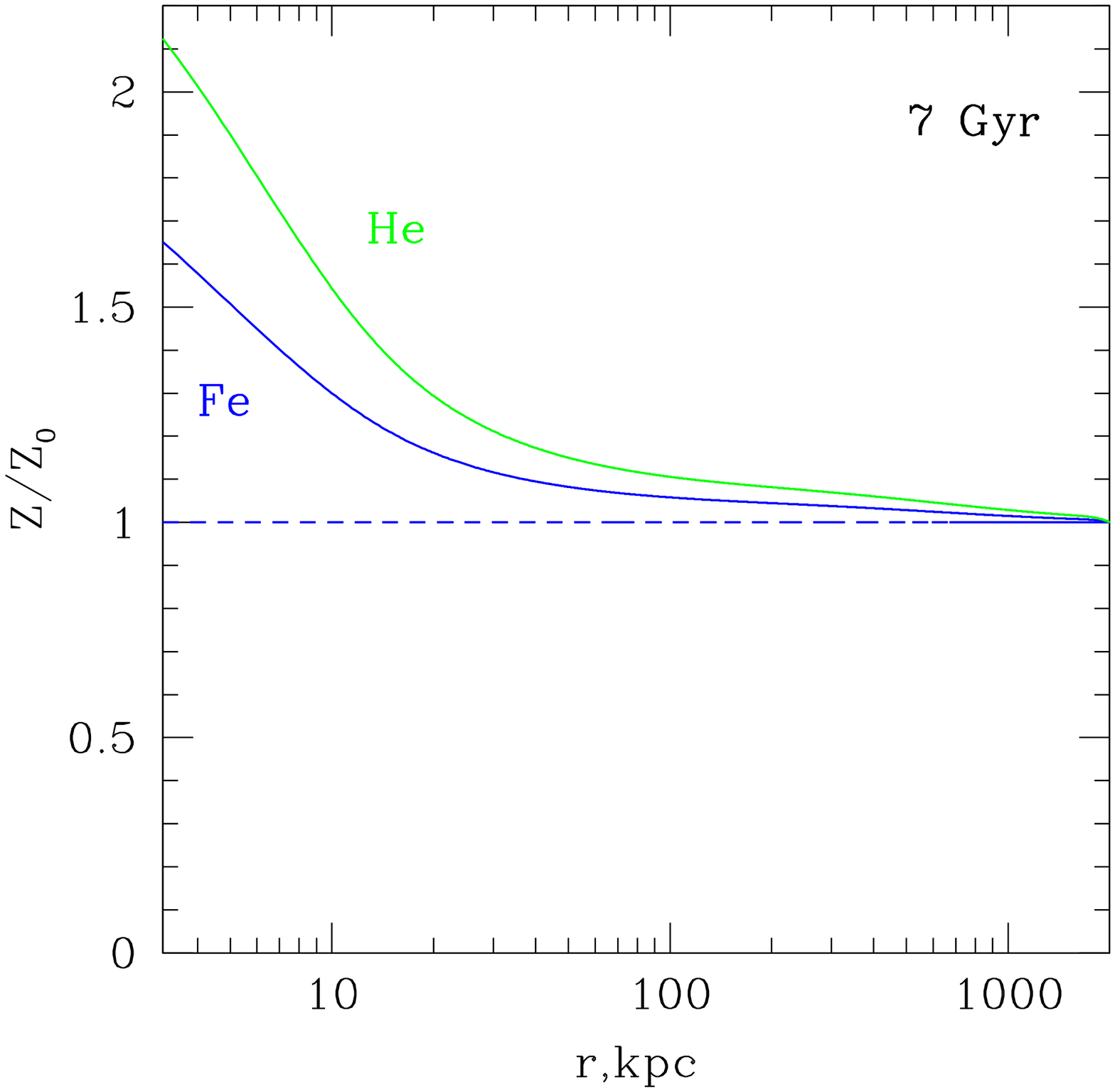}
}\end{centering}
\caption{
Same as in Fig.~\ref{fig:metprofiles}, but for an isothermal cluster from Fig.~\ref{fig:profiles2}
}
\label{fig:metprofiles2}
\end{figure*}

\section{Results and discussion}
\label{sec:results}

\subsection{Model cluster}
\label{sec:modelcluster}

For the temperature profile of the cluster,  $T(r)$,
we use the analytical best-fit model for A262 from 
\citet{vikhlinin06}.
Outside the range of its applicability the profile was extended as illustrated
 in the  Figure~\ref{fig:profiles}.
The total mass density of the cluster is assumed to follow the NFW profile
\citep{navarro97}:
\begin{equation}
\rho_{NFW}=\frac{\rho_s}{(r/r_s)(1+r/r_s)^2}.
\label{eq:rhodm}
\end{equation}
We take  $r_s=184$~kpc, the observed value for the  A262 
cluster \citep{vikhlinin06}. 

Given the $T(r)$ and $r_s$ and assuming hydrostatic equilibrium, 
we determine the value of $\rho_s$  resulting in gas density $\rho_g(r)$ 
which follows as close as possible
 the analytical best-fit model for gas density
from \citet{vikhlinin06}.
With the obtained value of normalisation, $\rho_s=1.03\times10^{-25}$~g/cm$^3$,
the total mass inside the radius $r_x=70$~kpc obtained integrating  the 
$\rho_{NFW}(r)$ equals the total mass calculated using the 
temperature and gas density profiles (and their extrapolations) from \citet{vikhlinin06},
assuming the hydrostatic equilibrium.
The overall agreement between these two masses is better than $\approx20-30\%$ in the 
range of 20--450~kpc with the limits reflecting the bounds of the region with a sufficient quality  
of data \citep{vikhlinin06}.
The final gas density profile is calculated using the adjusted NFW profile 
$\rho_{NFW}(r)$, temperature profile 
T(r) and assuming the hydrostatic equilibrium. 
The normalization of the gas density profile corresponds to a
gas mass fraction within $r_{500}$ (the radius where the overdensity
is equal to 500) of $f_{500}=0.14$.
The resulting total mass, gas density and temperature profiles are shown 
in  fig.~\ref{fig:profiles}.

To illustrate the role of the thermal diffusion, we also consider an isothermal 
cluster with temperature $kT=2$ keV and  the same total mass density profile. 
The initial  gas density profile is calculated similarly to above, assuming the 
hydrostatic equilibrium (fig.~\ref{fig:profiles2}).

We consider a 4-component plasma consisting of  H$^{+}$, He$^{+2}$, Fe$^{+26}$ 
and electrons.
The initial metallicity profiles are assumed to be flat with 
the mass fractions  of hydrogen and iron being 0.75 and
$1.8\cdot10^{-3}$ respectively.
To demonstrate the effects of diffusion for different species, we also 
perform calculations replacing the Fe$^{+26}$ by O$^{+8}$.

\subsection{Diffusion in clusters of galaxies: temperature vs. gravity}

We plot on fig.~\ref{fig:profiles} and \ref{fig:profiles2} the diffusion velocity
 profiles at the 
initial moment of time for a cool core and isothermal models respectively.
Positive values correspond to outflow of particles,  negative to inflow.
As is clear from the plots, the presence of temperature
 gradients drastically changes the character of diffusion.
While in an isothermal cluster the effect of diffusion is restricted to 
sedimentation of He and heavy nuclei, in the non-isothermal case
the picture  is more complicated. The combined effect of gravity and temperature gradients leads to formation of a complex diffusion velocity profile with prominent features at the positions of strong temperature gradients ($\sim$20~kpc and $\sim400$~kpc in fig.~\ref{fig:profiles}). 
At small radii, the sharp decrease of temperature  counteracts gravity and reverses the diffusion flow, evacuating heavy  elements from the cool core of the cluster. 
On the other hand, the negative temperature gradient in the outer part of the cluster works in the same direction as the force of gravity, increasing the velocity of gravitational sedimentation. 
Note that due to the lower temperatures and higher gas density, the diffusion velocities in the centers of cool core clusters are smaller than in higher temperature IGM usually considered in the context of the gravitational sedimentation of elements in clusters of galaxies \cite[e.g.][]{chuzhoy03}. The goal of the present analysis is to study the possible role of thermal diffusion in the IGM. We therefore focused on the cool core clusters having strong temperature gradients, rather than on the high temperature clusters characterized by higher diffusion velocities.
Because of the relatively small diffusion velocity,  ions can drift only through 
a small fraction of the cluster, $\la20$~kpc, during it's lifetime.

The overall effect of diffusion is further illustrated by the metallicity profiles after 3 and 7 Gyrs shown in 
figure~\ref{fig:metprofiles}.
At large radii the heavy elements flow inwards 
due to combined effect of thermal diffusion and gravity.
The temperature peak at $\sim300$~kpc leads to the enhancement
of heavy elements abundance at this radius.
At intermediate radii the temperature gradient is relatively weak 
and thus metals flow under the action of gravity 
to the center.
Due to the strong temperature gradient at $r\sim 20$~kpc,  the inward diffusion of 
heavy elements stops and the  
second peak is formed in heavy elements distribution. 
At the same time metals from the core of the cluster flow up 
the temperature gradient,  making the peak  more prominent and
 creating a depression in the metallicity at smaller radii. 
Because of the nearly flat temperature distribution in the inner $\sim 10$ kpc, assumed here, 
 the abundance at the center of the 
cluster is also somewhat enhanced by the gravitational sedimentation, with 
the effect being stronger for helium, 
than for iron.
In contrast to this, the picture is much simpler in the isothermal cluster model shown in fig.~\ref{fig:profiles2}. In this case the density profile evolution is driven by the gravitational sedimentation of heavy elements leading to a monotonic increase of their abundance towards the cluster center.

\begin{figure}
\hbox{
\includegraphics[width=0.5\textwidth]{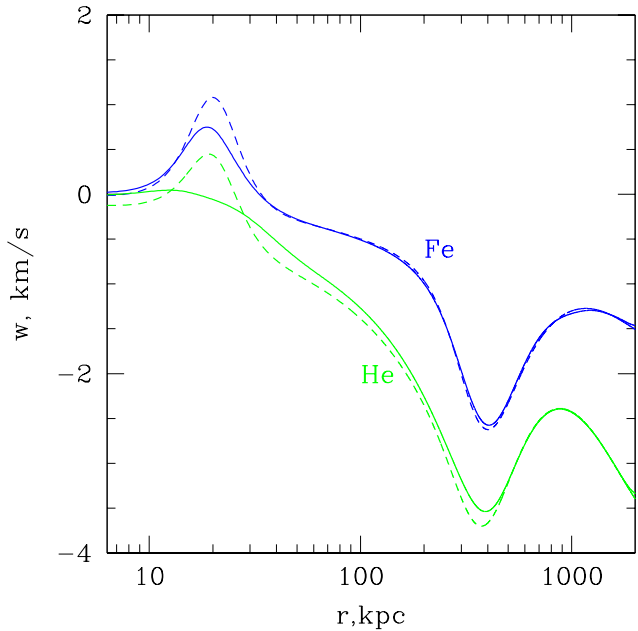}
}
\caption{
Diffusion velocity profiles for Fe (blue line) and He (green) in the model galaxy cluster after 
7 Gyr. Dashed lines show the initial diffusion velocities.
}
\label{fig:vdistrevol}
\end{figure}

In figure~\ref{fig:vdistrevol} we compare the diffusion velocity profiles 
after 7 Gyr with the initial distributions.
Clearly, the outflow of Fe at small radii is only slightly reduced, while for He 
it almost disappears. 
The diffusion velocities outside the inner parts of the cluster remain almost 
unchanged.

\subsection{Dependence on the atomic mass and charge}

Both gravitational sedimentation and
thermal diffusion velocities  depend on the ion species.
As is clear from Fig.~\ref{fig:profiles2}, the gravitational sedimentation in an 
isothermal cluster is the most 
effective for the helium ions, with the higher $Z$ ions such as Fe and O diffusing 
somewhat slower. 
At the same time, the diffusion velocities are almost equal for the latter two, 
$w_{Fe}\approx w_{O}$ \citep{chuzhoy03}.
For a non-isothermal temperature profile,
the equilibrium distribution of the heavy element 
depends on the element's charge with more highly 
charged species being more concentrated towards 
peaks in temperature profile of the cluster  (e.g. eq.~\ref{eq:equildistr}). 
As a result, the abundance profile of He 
appears to be smoother than that of  Fe (fig.~\ref{fig:metprofiles}).
On the other hand, on the timescales of interest, the effect of diffusion 
for elements heavier than He is identical from the observational point of view.  
The strong charge dependence of  thermal diffusion (eq.~\ref{eq:equildistr}) 
becomes relevant only on 
significantly longer timescales, due to the small $\nabla T$ in the 30-200~kpc 
interval.

\begin{figure*}
\begin{centering}\hbox{
\includegraphics[width=0.45\textwidth]{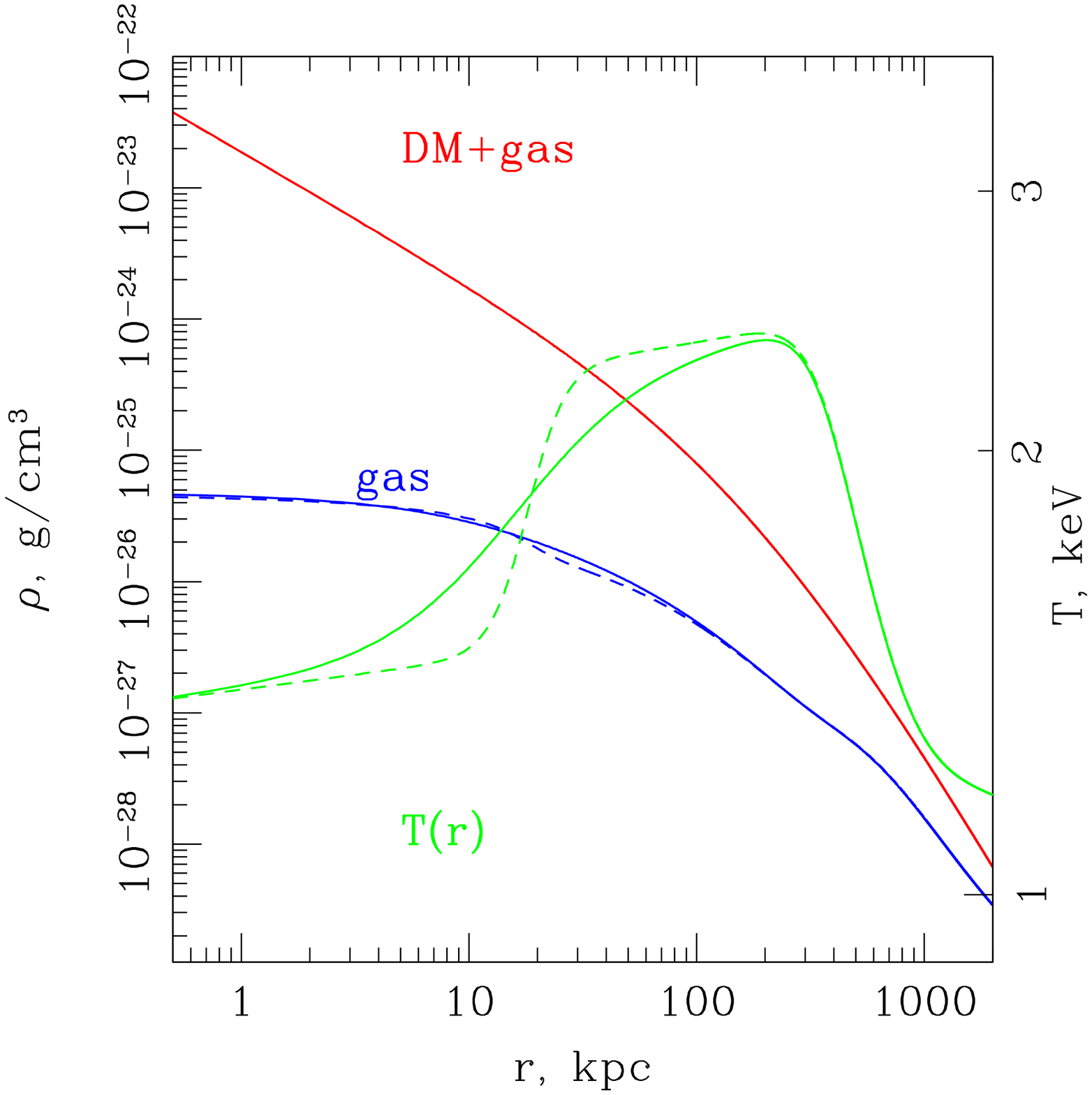}
\includegraphics[width=0.45\textwidth]{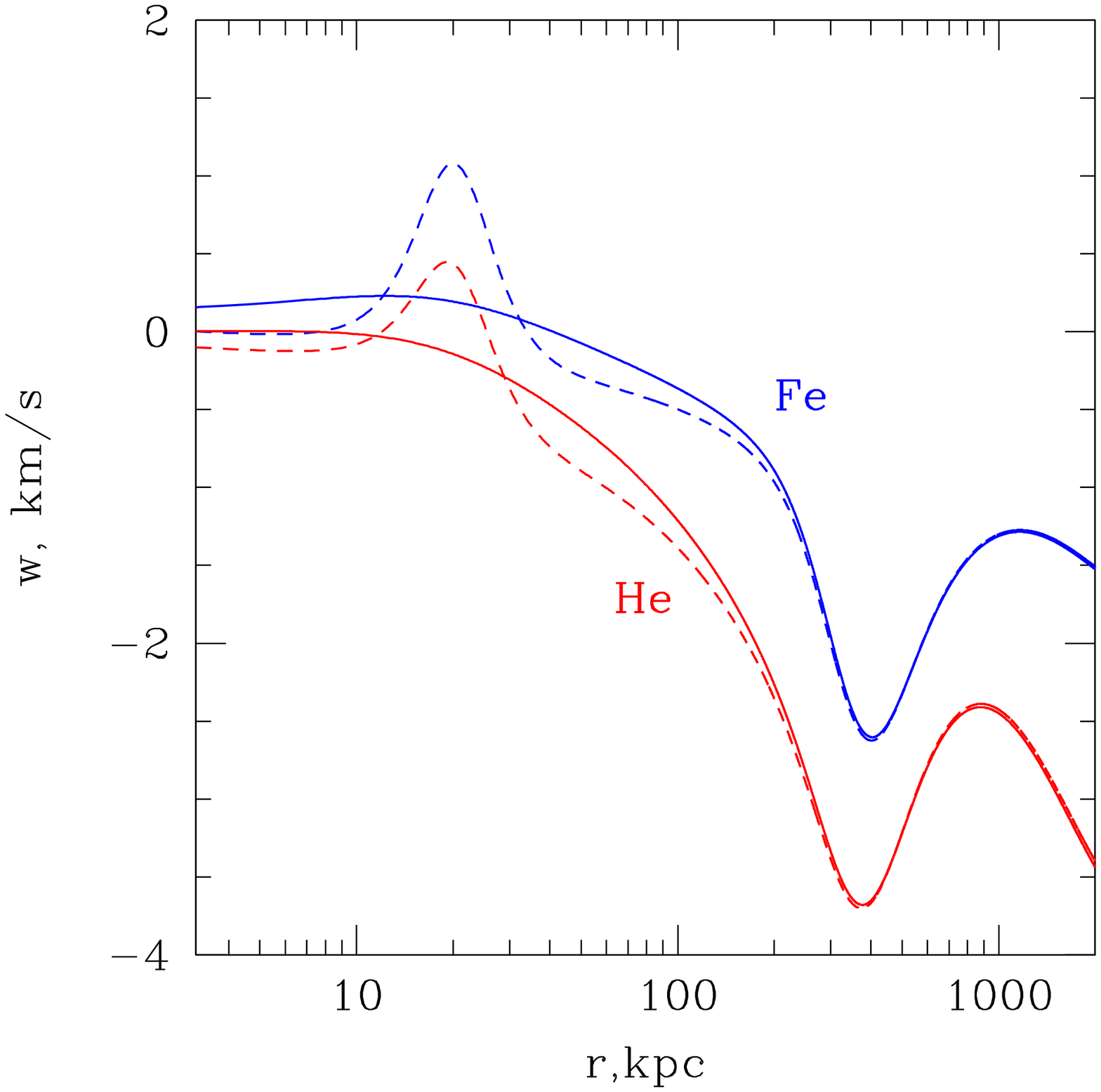}
}\end{centering}
\caption{{\em Left:} 
Dark Matter, gas 
density and temperature profiles in the model
galaxy clusters with an (ad hoc) temperature distribution with a moderate gradient in the core. 
The total mass distribution is the same as in our default model shown in fig.~\ref{fig:profiles}.
{\em Right:} Diffusion velocity profiles for He (red) and Fe (blue) at the initial moment of
 time. 
For comparison,  the dashed lines in both panels show profiles for the case of a large
 temperature gradient from fig.\ref{fig:profiles}.
Unlike the latter case, He flows inwards over all   radii.
}
\label{fig:diffvelocity2}
\end{figure*}

\begin{figure*}
\begin{centering}\hbox{
\includegraphics[width=0.45\textwidth]{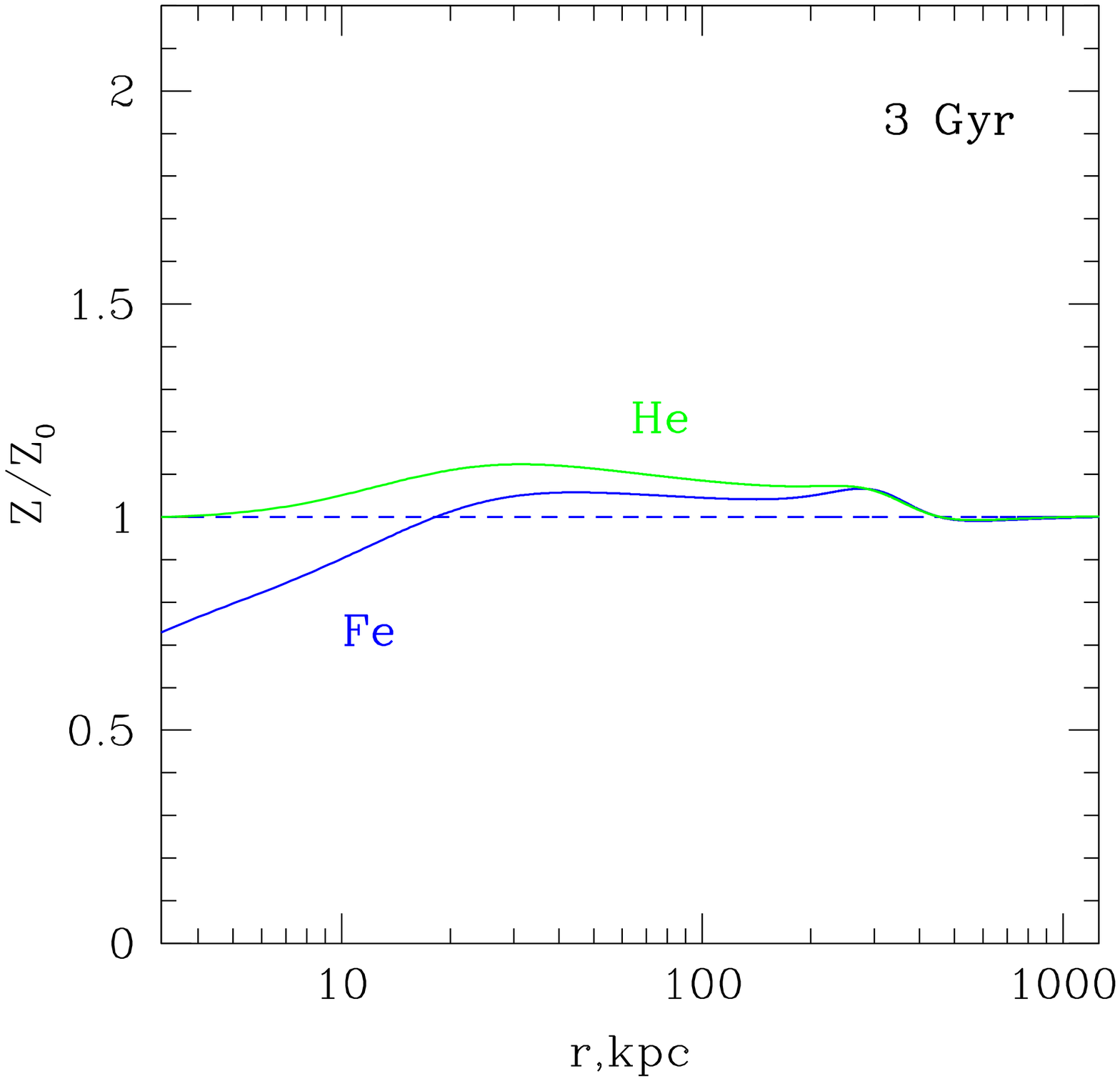}
\includegraphics[width=0.45\textwidth]{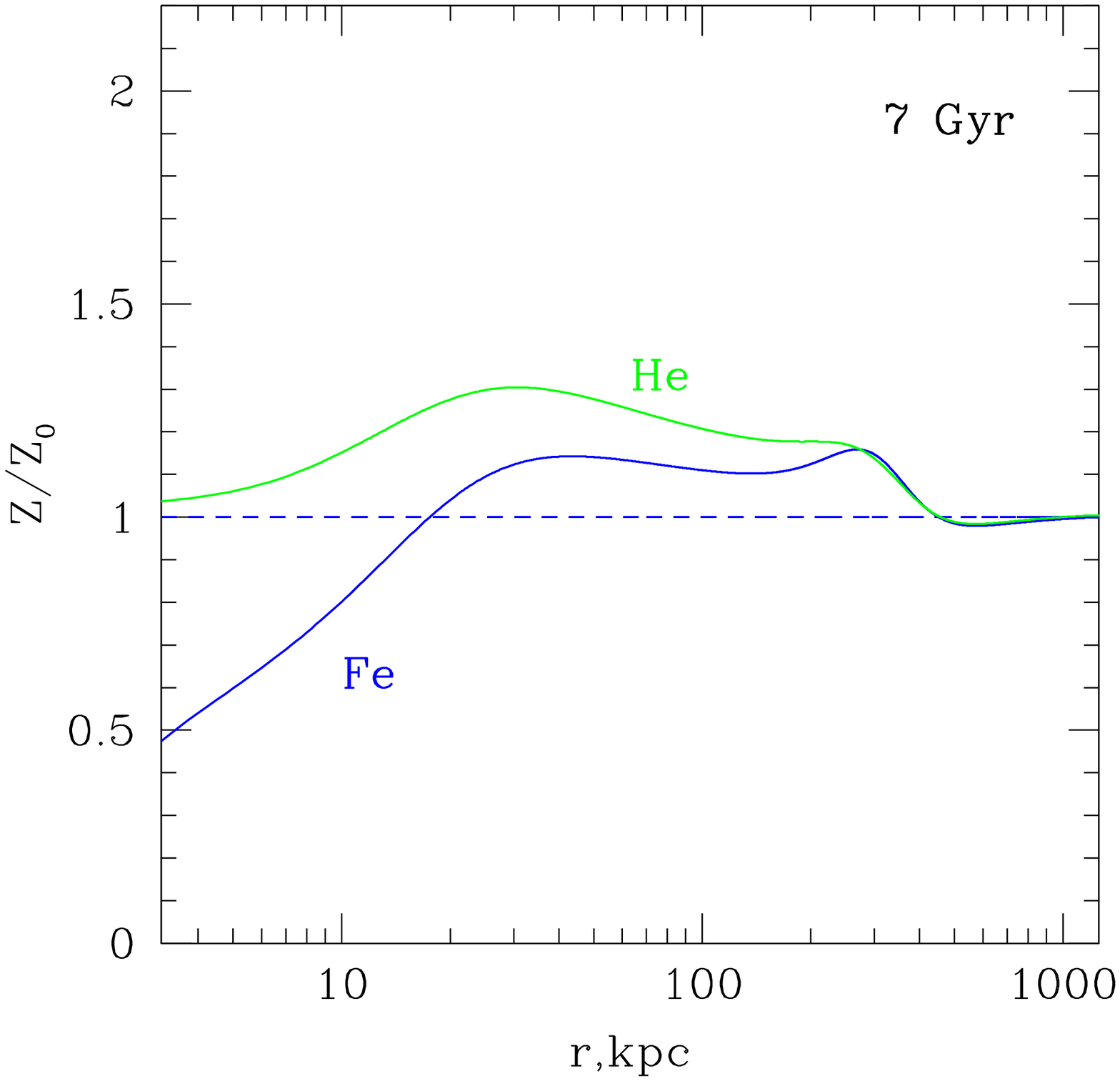}
}\end{centering}
\caption{{\em Left:} Abundance profiles for Fe (blue solid line), 
and He (green line) in the model  galaxy cluster from  fig.~\ref{fig:diffvelocity2}
after 3 Gyr. 
{\em Right:} The same, but after 7 Gyr. 
}
\label{fig:metprofiles3}
\end{figure*}

\subsection{Dependence on the cluster parameters}

The overall pattern of diffusion depends critically on the cluster parameters, 
due to the strong temperature, density and gravity dependence of 
diffusion velocities.
Since the typical temperatures in clusters of galaxies vary 
from  less than a 1~keV  up to $\sim10$~keV and the diffusion velocity 
scales with temperature as $w\propto T^{3/2}$ \citep[e.g.][]{thoul94}, 
the variations in the diffusion velocity may exceed an order of magnitude.
On the other hand, the temperature, gas density and gravity in clusters of galaxies
are by no means independent  \citep[e.g.][]{vikhlinin06}.
Thus the character of the diffusion is a result 
of the complex interplay between the cluster parameters.

The reversal of the diffusion flow for  He and metals  
at small radii (fig.~\ref{fig:profiles}) is a consequence of a  large  gradient in the
 temperature profile, which counteracts 
the force of gravity. The balance between these two forces is sensitive to the 
details of the temperature distribution. Rather small and otherwise almost insignificant
 modifications of the temperature profile may lead to the diffusion pattern
 different from that shown in
 fig.~\ref{fig:profiles} and \ref{fig:metprofiles}.
To illustrate this, we consider a cluster of galaxies with an (ad hoc) smoother temperature
 profile, characterized by smaller gradients  in the cluster core, but with the same total mass distribution.
 The cluster parameters  are shown in fig.~\ref{fig:diffvelocity2}.
 As is obvious from the 
figure, the initial distribution of the gas density does not differ significantly from our
 default model.
However the diffusion velocity profile changes dramatically -- the positive velocity peaks  disappear for both elements and He velocity becomes negative in the entire range of radii (fig.~\ref{fig:diffvelocity2}).
The resulting metallicity profile is correspondingly significantly smoother 
(cf. fig.~\ref{fig:metprofiles}). However, thermal diffusion counteracts
the effect of the gravitational sedimentation in this case as well, significantly 
reducing its effect for He and resulting in a remarkable central depression in the Fe
abundance.  

Further decrease of the temperature gradient in the cluster core 
can make gravity the dominant factor.
Based on the solution of Burgers' equations we 
estimate that for the gravitational potential considered here, 
the thermal diffusion equals to the force of gravity for the temperature gradient of
$\Delta T\sim 0.1$ keV per 20~kpc. Note that this is $\sim$2 times smaller than the temperature gradient in the profile shown  in the fig.~\ref{fig:diffvelocity2},  $\Delta T\sim 0.2$ keV per 20~kpc at  $r=20$~kpc.

\subsection{Turbulent mixing and metals injection from cD galaxy}

\label{sec:mixing}

An idealized model constructed in this paper 
 ignores turbulent mixing and metal enrichment of the IGM by the central cD galaxy - the two important factors which potentially may have a significant impact on the metallicity profiles.
The turbulent mixing tends to smoothen the abundance distributions, while the cD galaxy increases the heavy element  abundances in the inner parts of the cluster due to metals produced by  (mostly)  SNIa explosions.
These two factors are discussed below.

In order to estimate role of the cD galaxy, we 
measure the K-band luminosity of A262 using the image from 2MASS Extended Source Image Server 
(http://irsa.ipac.caltech.edu/applications/2MASS/ PubGalPS/)
and based on the calibration of \citet{mannucci05} 
for E/S0 galaxies convert it to the SN Ia rate.
Combining the latter with the iron yield of $\eta\sim0.7$~M$_{\odot}$ \citep{iwamoto99} and assuming perfect mixing of the SNIa ejecta with IGM,
we obtain the iron injection rate of $8.0\cdot10^{-3}$~M$_{\odot}$/yr
within  the inner $\approx19$~kpc of the cluster.
On the other hand, given the  diffusion velocity, $w_{\rm Fe}\approx10^5$~cm/s, and
concentration of iron  at this radius  (see fig.~\ref{fig:profiles}), we obtain 
the diffusion-driven iron outflow from the same region of
$\approx2.3\cdot10^{-3}$~M$_{\odot}$/yr, which is several times
smaller than the iron injection rate.
The relation between the outflow and injection rates  is opposite for $\alpha$-elements.
Indeed, assuming the oxygen yield of $\approx0.1$~M$_{\odot}$ and solar
metallicity, the oxygen injection rate is 
$\approx1.1\cdot10^{-3}$~M$_{\odot}$/yr, which is several times smaller than the 
outflow rate, $\approx4.9\cdot10^{-2}$~M$_{\odot}$/yr.
Taking these numbers at the face value, we may conclude that the impact of the cD galaxy on metallicity
profiles  varies from significant for iron peak elements to negligible
for $\alpha$-elements.

\begin{figure}
\hbox{
\includegraphics[width=0.5\textwidth]{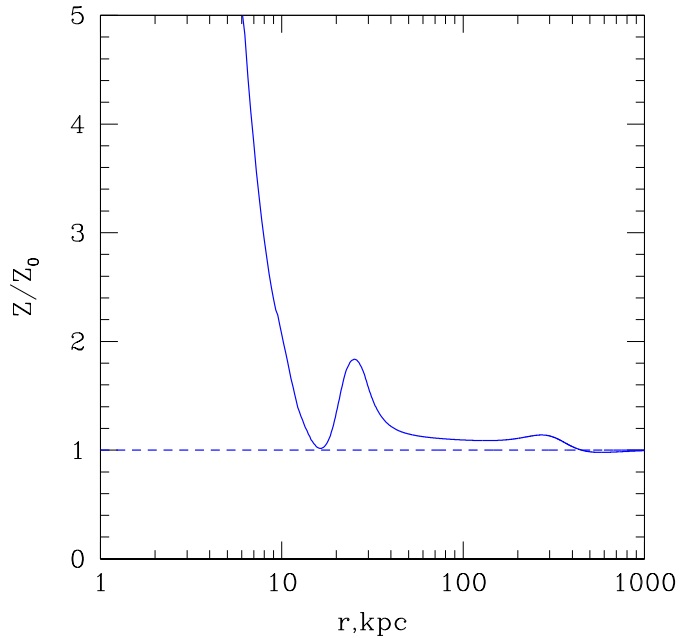}
}
\caption{ Abundance profile for Fe in the model  galaxy cluster 
with injection of metals by cD galaxy after 7 Gyr. See section \ref{sec:mixing} for details.}
\label{fig:metprofiles4}
\end{figure}

We assumed above that complete mixing of the Supernova ejecta with IGM takes place.
However, the details of the SN ejecta interaction with the ICM are poorly understood. Indeed, the assumption of the total mixing leads to unreasonably high iron abundance in the cluster core, as illustrated by the following calculation.
Assuming that the iron injection rate in A262 follows the stellar mass
profile and taking into account 
the metal injection by Supernovae via an additional source term in
eq.~\ref{eq:continuity}, 
we repeated the calculation of the metallicity profile
evolution described in sec.~\ref{sec:experiment}.\footnote{Note, that due to
the limited coverage of A262 by 2MASS, the metal injection rate at
$r\ga19$~kpc is somewhat underestimated. This, however, does not
change our conclusions.}
This resulted in a strongly peaked iron profile (Fig.\ref{fig:metprofiles4}) with unrealistically
high metallicity  in the  center of the cluster, $Z\approx7~Z_{\odot}$ within the inner 10 kpc.
Such a high iron abundance in the cluster core contradicts to observations of clusters of galaxies, in particular to the iron distribution in A262 measured by 
\cite{vikhlinin05}, pointing at the need of a more sophisticated
model.
Similar results were obtained by \citet{rebusco06}, who found the
observed metallicity profiles in galaxy clusters to be broader than 
predicted under the assumption  that the rate of metals
injection follows stellar light distribution.
A possible explanation of such a discrepancy is
offered by the turbulent mixing driven by the central AGN activity. This requires 
turbulent diffusion coefficients   of $6\cdot10^{27}-5\cdot10^{29}$~cm$^2$/s, corresponding to mixing length
scales of $1-35$~kpc and gas velocities $\approx200-400$~km/s \citep{rebusco06}.
However,  the role of this process in the particular case of A262 and other cool core clusters is unclear. Indeed, the presence of  steep temperature and abundance gradients
\citep{vikhlinin05,vikhlinin06}  rules out any significant mixing operating on
spatial scales exceeding $\sim10$~kpc, thus excluding a large part of the parameters space required for the turbulent mixing interpretation.

To conclude, the role of mixing and metal enrichments remains to be controversial. An attempt to include them in the model in a straightforward manner leads to an apparent conflict with observations. This suggests that the role of these processes may be overestimated. Secondly, this points at the need of more sophisticated models. Consideration of such models will be a subject of the forthcoming paper.

\section{Conclusions}
\label{sec:discussion}

We considered the effects  of element diffusion in the intergalactic medium 
on the metallicity profiles in clusters of galaxies.
The diffusion processes in the IGM are driven by the gravity, concentration and 
temperature gradients.
Gravity forces metals to sink towards the center, concentration
gradients act to restore the
uniform distribution, and  the temperature gradients tend to evacuate
heavy elements from the low temperature regions in the IGM.
So far, attention has been paid to gravitational sedimentation, 
while thermal diffusion has been largely excluded from  consideration.
We demonstrate that the latter may drastically change the 
pattern of the diffusion and thus may play a significant role in  formation of the 
metallicity distribution in IGM.

We performed numerical simulations of the diffusion in the IGM of clusters of galaxies based on the
 full set of Burgers' equations. In order to illustrate the effect of thermal diffusion 
we considered three different models of cluster of galaxies with the same total mass 
profile but different temperature distributions: isothermal, and with large and moderate 
temperature gradient (figs.~\ref{fig:profiles}, \ref{fig:profiles2}, \ref{fig:diffvelocity2}).
The  evolution of metallicity profiles are shown on figs.~\ref{fig:metprofiles}, 
\ref{fig:metprofiles2} and \ref{fig:metprofiles3}.
As illustrated by these plots, the combined effects of the temperature gradients and 
gravity may result in a rather complex metallicity profiles in clusters of galaxies. 

The efficiency of transport processes in IGM is still under debate. 
However, if element diffusion does operate in clusters of galaxies, its overall 
effect on metallicity distributions may be opposite to what was previously thought,
 based on simplified considerations of isothermal IGM. Our results show that even 
moderate temperature gradients, $\sim  0.5$ keV per 100~kpc may be 
sufficient to counteract  the force of gravity and to stop and even to reverse the 
diffusion flow of metals, thus preventing enhancement of heavy element abundances 
in the center due to gravitational sedimentation.
It is especially relevant to the cool core clusters, where 
the thermal diffusion may  lead to depletion of metals in the cluster core.
The implications of this process for the diagnostics of the low temperature plasma  in
 the centers of clusters of galaxies will be considered in the forthcoming paper 
(Shtykovskiy \& Gilfanov, in preparation).

{\em Acknowledgements.}
PS acknowledges support from the President of the Russian Federation grant
NSh-5579.2008.2, the program of the Presidium of Russian Academy of
Sciences (RAS) ''The origin, structure and evolution
of objects in the Universe'' (P-07), and the program of RAS OFN-17. 
PS  would like to thank Max-Planck-Institute for Astrophysics, where part of this work was 
done for their hospitality.
We also would like to thank the anonymous referee for useful comments on
the manuscript.

\label{lastpage}

\end{document}